\documentstyle[11pt]{article}
\newcommand{\be}{\begin{equation}}
\newcommand{\ee}{\end{equation}}
\newcommand{\bea}{\begin{array}}
\newcommand{\eea}{\end{array}}

\title{SOME KERNELS ON A RIEMANN SURFACE}
\author{ROBERT CARROLL\\UNIVERSITY OF ILLINOIS\\URBANA, IL 61801\thanks
{email: rcarroll@symcom.math.uiuc.edu}}
\date{December, 1996}

\begin{document}

\bibliographystyle{plain}
\maketitle

\begin{abstract}
We discuss certain kernels on a Riemann surface constructed mainly via 
Baker-Akhiezer (BA) functions and indicate relations to dispersionless
theory.

\end{abstract}

\section{INTRODUCTION}
\renewcommand{\theequation}{1.\arabic{equation}}\setcounter{equation}{0}

A preliminary sketch of some of this material was given in \cite{cn}, and
in view of recent developments in \cite{co,cp,cs,dq,kt,kf,mz,ne}, we have
expanded and rewritten that manuscript into \cite{cz} plus the present
paper.  We develop here, on a Riemann surface, a version of a kernel
constructed in \cite{ch} for dispersionless KP (dKP), which is in fact
equivalent to the dispersionless differential Fay identity (cf.
also \cite{kn,km,ta} for related material).  There is also some discussion
of other kernels and relations among them.

\section{BACKGROUND}
\renewcommand{\theequation}{2.\arabic{equation}}\setcounter{equation}{0}

\subsection{BA functions}

For completeness we recall certain information about Riemann surfaces
and BA functions following \cite{cn,cz}.  The use of BA functions goes
back to \cite{bc} for example but was developed in modern times by the
Russian school (see e.g. \cite{bb,ca,cm,da,dc,dd,de,ka,ke,kj,nd,tb}).  
Thus take an arbitrary
Riemann surface $\Sigma$
of genus $g$, pick a point $Q\sim P_{\infty}$ and a 
local variable $1/k$ near $Q$
such that $k(Q) = \infty$, and, for illustration, take $q(k)=kx+k^2y+k^3t$. 
Let $D = P_1 + \cdots + P_g$ be a non-special
divisor of degree $g$ and write $\psi$ for the (unique up to a constant
multiplier by virtue of the Riemann-Roch theorem)
Baker-Akhiezer (BA) function characterized by the properties
({\bf A}) $\psi$ is meromorphic on $\Sigma$ except for $Q$ where $\psi
(P)exp(-q(k))$ is analytic 
and (*) $\psi\sim exp(q(k))[1 + \sum_1^{\infty}(\xi_j/
k^j)]$ near $Q$.  ({\bf B}) On $\Sigma/Q,\,\,\psi$ has only a finite
number of poles (at the $P_i$).
In fact $\psi$ can be taken in the form ($P\in\Sigma,\,\,
P_0\not= Q$)
\be
\psi(x,y,t,P) =  exp[\int^P_{P_0}(xd\Omega^1 + yd\Omega^2 + td\Omega^3)]
\cdot\frac{\Theta({\cal A}(P) + xU + yV + tW + z_0)}{\Theta({\cal A}
(P) + z_0)}
\label{psi}
\ee
where $d\Omega^1 = dk + \cdots,\,\,d\Omega^2 = d(k^2) + \cdots,\,\,
d\Omega^3 = d(k^3) + \cdots, U_j = \int_{B_j}d\Omega^1,\,\,V_j = \int_
{B_j}d\Omega^2,\,\,W_j = \int_{B_j}d\Omega^3\,\,(j = 1,\cdots,g),\,\,z_0
= -{\cal A}(D) - K$, and $\Theta$ is the Riemann theta function.
The symbol $\sim$ will be used generally to mean ``corresponds
to" or ``is associated with"; occasionally it also denotes asymptotic
behavior and this should be clear from the context.
Here the
$d\Omega_j$ are meromorphic differentials of second kind normalized via
$\int_{A_k}d\Omega_j = 0\,\,(A_j,\,B_j$ are canonical homology cycles)
and we note that $xd\Omega^1 + yd\Omega^2 + td\Omega^3\sim
dq(k)$ normalized.
${\cal A}$ is the Abel-Jacobi map ${\cal A}(P) = (\int^P_{P_0}d\omega_k)$,
where
the $d\omega_k$ are normalized holomorphic differentials, $k = 1,\cdots,g,
\,\,\int_{A_j}d\omega_k = \delta_{jk}$, and $K = (K_j)\sim$ Riemann
constants ($2K = -{\cal A}(K_{\Sigma})$ where $K_{\Sigma}$ is the
canonical class of $\Sigma\sim$ equivalence class of meromorphic 
differentials). Thus $\Theta({\cal A}(P) + z_0)$ has exactly $g$ zeros
(or vanishes identically.  The paths of integration are to be the
same in computing $\int_{P_0}^Pd\Omega^i$ or ${\cal A}(P)$ and it is
shown in \cite{bc,bb,cn,dc} that $\psi$ is well defined (i.e. path independent).  
Then the $\xi_j$ in (*) can be computed
formally and one determines Lax operators $L$ and $A$ such that
$\partial_y\psi = L\psi$ with $\partial_t\psi = A\psi$.  Indeed, given
the $\xi_j$ write $u = -2\partial_x\xi_1$ with $w = 3\xi_1\partial_x\xi_1
-3\partial^2_x\xi_1 - 3\partial_x\xi_2$.  Then formally, near $Q$, one
has $(-\partial_y + \partial_x^2 + u)\psi = O(1/k)exp(q)$ and 
$(-\partial_t + \partial^3_x + (3/2)u\partial_x + w)\psi = O(1/k)exp(q)$
(i.e. this choice of $u,\,w$ makes the coefficients of $k^nexp(q)$ vanish
for $n = 0,1,2,3$).  Now define $L = \partial_x^2 + u$ and $A = \partial^3_x
+ (3/2)u\partial_x + w$ so $\partial_y\psi = L\psi$ and $\partial_t\psi
= A\psi$.  This follows from the uniqueness of BA functions with the same
essential singularity and pole divisors (Riemann-Roch). 
Then we have, via compatibility
$L_t - A_y = [A,L]$, a KP equation $(3/4)u_{yy} = \partial_x[u_t
-(1/4)(6uu_x + u_{xxx})]$ and therefore such KP equations are parametrized
by nonspecial divisors or equivalently by points in general position
on the Jacobian variety $J(\Sigma)$.
The flow variables $x,y,t$ are put in by hand in ({\bf A}) via
$q(k)$ and then miraculously reappear in the theta function via
$xU+yV+tW$; thus the Riemann surface itself contributes to establish
these as linear flow variables on the Jacobian.
The pole positions
$P_i$ do not vary with $x,y,t$ and $(\dagger)\,\,
u = 2\partial^2_x log\Theta(xU + yV  + tW + z_0) + c$ 
exhibits $\Theta$ as a tau function.
\\[3mm]\indent
We recall also that
a divisor $D^{*}$ of degree $g$ is dual to $D$ (relative to $Q$) if
$D + D^{*}$ is the null divisor of a meromorphic differential $d\hat{\Omega}
= dk + (\beta/k^2)dk + \cdots$ with a double pole at $Q$ (look at
$\zeta = 1/k$ to recognize the double pole).  Thus 
$D + D^{*} -2Q\sim K_{\Sigma}$ so ${\cal A}(D^{*}) - {\cal A}(Q) + K =
-[{\cal A}(D) - {\cal A}(Q) + K]$.  One can define then a function
$\psi^{*}(x,y,t,P) = exp(-kx-k^2y-k^3t)[1 + \xi_1^{*}/k) + \cdots]$
based on $D^{*}$ (dual BA function)
and a differential $d\hat{\Omega}$ with zero divisor $D+D^*$, such that
$\phi = \psi\psi^{*}d\hat{\Omega}$ is
meromorphic, having for poles only
a double pole at $Q$ (the zeros of $d\hat{\Omega}$ cancel
the poles of $\psi\psi^{*}$).  Thus $\psi\psi^*d\hat{\Omega}\sim \psi\psi^*(1+
(\beta/k^2+\cdots)dk$ is meromorphic with a second order pole at $\infty$,
and no other poles.  
For $L^{*} = L$ and $A^{*} = -A + 2w
-(3/2)u_x$ one has then $(\partial_y + L^{*})\psi^{*} = 0$ and
$(\partial_t + A^{*})\psi^{*} = 0$.  Note that  
the prescription above seems to specify for $\psi^*$
($\vec{U}=xU+yV+tW,\,\,z_0^* =
-{\cal A}(D^*)-K$)
\be
\psi^*\sim e^{-\int^P_{P_o}(xd\Omega^1+yd\Omega^2+td\Omega^3)}
\cdot\frac{\Theta({\cal A}(P)-\vec{U}+z_0^*)}
{\Theta({\cal A}(P)+z_0^*)}
\label{star}
\ee
\indent
In any event the message here is that for any Riemann surface 
$\Sigma$ one can
produce a BA function $\psi$ with assigned flow variables $x,y,t,\cdots$
and this $\psi$ gives rise to a 
(nonlinear) KP equation with solution $u$ linearized
on the Jacobian $J(\Sigma)$.
For averaging with KP (cf. \cite{cn,cz,fb,ka}) 
we can use formulas (cf. (\ref{psi}) and
(\ref{star}))
\be
\psi = e^{px+Ey+\Omega t}\cdot\phi(Ux+Vy
+Wt,P)
\label{YDD}
\ee
\be
\psi^* = e^{-px-Ey-\Omega t}\cdot\phi^*
(-Ux-Vy-Wt,P)
\label{YEE}
\ee
to isolate the quantities of interest in averaging
(here $p=p(P),\,\,E = E(P),\,\,\Omega =
\Omega(P),$ etc.)
We think here of a general Riemann surface $\Sigma_g$ with holomorphic
differentials $d\omega_k$ and quasi-momenta and quasi-energies
of the form $dp=d\Omega^1,\,\,dE=d\Omega^2,\,\,d\Omega=d\Omega^3,\cdots
\,\,(p=\int_{P_0}^Pd\Omega^1$ etc.) where the $d\Omega^j=d\Omega_j=
d(\lambda^j+O(\lambda^{-1}))$ are meromorphic differentials of the second
kind.  Following \cite{ka} one could
normalize now via $\Re\int_{A_j}d\Omega^k=
\Re\int_{B_j}d\Omega^k=0$.  Then write e.g. 
$U_k=(1/2\pi i)\oint_{A_k}dp$ and $U_{k+g}=-(1/2\pi i)\oint_{B_k}dp\,\,
(k=1,\cdots,g)$ with similar stipulations for $V_k\sim\oint d\Omega^2,\,\,
W_k\sim\oint d\Omega^3,$ etc.  This leads to real $2g$ period vectors
and evidently one could also normalize via $\oint_{A_m}d\Omega^k=0$ or
$\Im\oint_{A_m}d\Omega^k=\Im\oint_{B_m}d\Omega^k=0$ (further we set
$B_{jk}=\oint_{B_k}d\omega_j$).

\subsection{KP and dKP}

We follow here \cite{ch,cl} (cf. also \cite{kn,ta}) and begin
with two pseudodifferential operators ($\partial = 
\partial/\partial x$),
\be
L = \partial + \sum_1^{\infty}u_{n+1}\partial^{-n};\, \ \,W = 1 + 
\sum_1^ {\infty}w_n\partial^{-n} \ ,
\label{ZQ}
\ee
called the Lax operator and gauge operator respectively, where 
$L = W\partial\,W^{-1}$. The KP hierarchy 
then is determined by
the Lax equations ($\partial_n = \partial/\partial t_n$), \be \partial_n L = 
[B_n,L] = B_n L - L B_n \ , 
\label{AAC} \ee 
where $B_n = L^n_{+} $ is the 
differential part of $L^n = L^n_{+} + L^n_{-} = \sum_0^{\infty}
\ell_i^n\partial^i + \sum_{-\infty}^ {-1}\ell_i^n\partial^i$. One can also 
express this via the Sato equation $\partial_n W\,W^{-1} = -L^n_{-}$ 
which is particularly well adapted to the dKP theory. Now define the wave 
function
via
\be
\psi = W\,e^{\xi} = w(t,\lambda)e^{\xi};\, \ \,\xi = \sum_1^{\infty}t_n 
\lambda^n;
\, \ \,w(t,\lambda) = 1 + \sum_1^{\infty}w_n(t)\lambda^{-n} \ , 
\label{AAE} 
\ee 
where $t_1 = x$. There is also an adjoint wave function $\psi^{*} = 
W^{*-1} \exp(-\xi) = w^{*}(t,\lambda)\exp(-\xi),$ with $w^{*}(t,\lambda) = 1 + 
\sum_1^ {\infty}w_i^{*}(t)\lambda^{-i}$ and one has equations 
$L\psi = 
\lambda\psi;\, \ \,\partial_n\psi = B_n\psi;\, \ \,L^{*}\psi^{*} = 
\lambda\psi^{*};\, \ \,\partial_n\psi^{*} = -B_n^{*}\psi^{*}$.
Note that the KP hierarchy (\ref{AAC}) is then given by the compatibility 
conditions among these equations, treating $\lambda$ as a constant.
Next one 
has the fundamental tau function $\tau(t)$ and vertex operators ${\bf X},
\,\,{\bf X}^{*}$ satisfying
\be
\psi(t,\lambda) = \frac{{\bf X}(\lambda)\tau (t)}{\tau (t)} = \frac{e^{\xi} 
G_{-}(\lambda)\tau (t)}{\tau (t)} = \frac{e^{\xi}\tau(t-[\lambda^{-1}])} 
{\tau (t)};
\label{AAG}
\ee
$$\psi^{*}(t,\lambda) = \frac{{\bf X}^{*}(\lambda)\tau (t)} {\tau (t)} = 
\frac{e^{-\xi}
G_{+}(\lambda)\tau (t)}{\tau (t)} = \frac{e^{-\xi}\tau(t+[\lambda^{-1}])} 
{\tau (t)} $$
where $G_{\pm}(\lambda) = \exp(\pm\xi(\tilde{\partial},\lambda^{-1}))$ with 
$\tilde{\partial} = (\partial_1,(1/2)\partial_2,(1/3)\partial_3, \cdots)$ 
and $t\pm[\lambda^{-1}]= (t_1\pm \lambda^{-1},t_2 \pm (1/2) \lambda^{-2}, 
\cdots)$.
One writes also
\be
e^{\xi} = \exp \left({\sum_1^{\infty}t_n\lambda^n}\right) = \sum_0^ 
{\infty}\chi_j(t_1, t_2, \cdots ,t_j) \lambda^j  \label{AAH} \ee 
where 
the $\chi_j$ are the elementary Schur polynomials, which arise in many 
important formulas (cf. below). 
\\[3mm]\indent We mention now the famous 
bilinear identity which generates the entire KP hierarchy. This has the 
form $({\bf H})\,\,
\oint_{\infty}\psi(t,\lambda)\psi^{*}(t',\lambda)d\lambda = 0$ 
where $\oint_{\infty}(\cdot)d\lambda$ is the residue integral about 
$\infty$, which we also denote $Res_{\lambda}[(\cdot)d\lambda]$. Using 
(\ref{AAG}) this can also be written in terms of tau functions as \be 
\oint_{\infty}\tau(t-[\lambda^{-1}])\tau(t'+[\lambda^{-1}]) 
e^{\xi(t,\lambda)-\xi(t',\lambda)}d\lambda = 0  \label{AAJ} 
\ee This leads to the characterization of the tau function in 
bilinear form expressed via ($t\to t-y,\,\,t'\to t+y$) 
\be 
\left(\sum_0^{\infty}\chi_n(-2y)\chi_{n+1}(\tilde{\partial})
e^{\sum_1^{\infty}
y_i \partial_i}\right)\tau\,\cdot\,\tau = 0  \label{AAK} 
\ee 
where $\partial^m_j a\,\cdot\,b = (\partial^m/
\partial s_j^m) a(t_j+s_j)b(t_j-s_j)|_{s=0}$ and $\tilde{\partial} = 
(\partial_1,(1/2) \partial_2,(1/3)\partial_3,\cdots)$. 
In particular, we have from the 
coefficients of $y_n$ in (\ref{AAK}), $({\bf HE})\,\,
\partial_1\partial_n\tau \cdot \tau = 2 \chi_{n+1} (\tilde{\partial}) 
\tau \cdot \tau$, 
which are called the Hirota bilinear equations.  One has also 
the Fay identity via (cf.
\cite{ab,ch} - c.p. means cyclic permutations) \be \sum_{c.p.}
(s_0-s_1)(s_2-s_3)\tau(t+[s_0]+[s_1])\tau(t+[s_2]+[s_3]) = 0  \label{ZR}
\ee
which can be derived from the bilinear identity (\ref{AAJ}). 
Differentiating this in $s_0$, then setting $s_0 = s_3 = 0$, then 
dividing by $s_1 s_2$, and finally shifting $t\to t-[s_2]$, leads to 
the differential Fay identity,
\begin{eqnarray}
\nonumber
& &\tau(t)\partial\tau(t+[s_1]-[s_2]) - \tau(t+[s_1] -[s_2])\partial 
\tau(t) \\ & &= (s_1^{-1}-s_2^{-1}) \left[\tau(t+[s_1]-[s_2]) \tau(t) - 
\tau(t+[s_1])\tau(t-[s_2])\right]  \label{AAL} \end{eqnarray} 
The 
Hirota equations
after (\ref{AAK}) can be also derived from (\ref{AAL}) by taking the 
limit $s_1 \to s_2$. The identity (\ref{AAL}) will play an important 
role later. \\[3mm]\indent
Now for the dispersionless theory (dKP) one can think of fast and slow 
variables, etc., or averaging procedures, but simply one takes 
$t_n\to\epsilon t_n = T_n\,\,(t_1 = x\to \epsilon x = X)$ in the 
KP equation $u_t = (1/4)u_{xxx} + 3uu_x + (3/4)\partial^{-1}u_{yy},
\,\, (y=t_2,\,\,t=t_3)$, with $\partial_n\to \epsilon\partial/\partial 
T_n$ and $u(t_n)\to U(T_n)$ to obtain $\partial_T U = 3UU_X + 
(3/4)\partial^ {-1}U_{YY}$ when $\epsilon\to 0\,\,(\partial =
\partial/\partial X$ now). Thus the dispersion term $u_{xxx}$ is
removed. In terms of hierarchies we write $({\bf L}):\,\,
L_{\epsilon} = \epsilon\partial + \sum_1^{\infty}u_{n+1}(T/\epsilon) 
(\epsilon\partial)^{-n}$ 
and think of $u_n( T/\epsilon)= U_n(T) + O(\epsilon)$, etc. One takes 
then a WKB form for the wave function with the action $S$  
\be \psi = \exp \left[\frac{1}{\epsilon}S(T,\lambda) \right] 
\label{AAN} 
\ee 
Replacing now $\partial_n$ by $\epsilon\partial_n$, where $\partial_n = 
\partial/\partial T_n$ now, we define $P = \partial S = S_X$. Then 
$\epsilon^i\partial^i\psi\to P^i\psi$ as $\epsilon\to 0$ and the equation 
$L\psi = \lambda\psi$ becomes $(\bullet\heartsuit\bullet)\,\,
\lambda = P + \sum_1^{\infty}U_{n+1}P^{-n};\, \ \,P = \lambda - 
\sum_1^{\infty}P_{i+1}\lambda^{-i}$, 
where the second 
equation is simply the inversion of the first. We also note from 
$\partial_n\psi =
B_n\psi = \sum_0^nb_{nm}(\epsilon\partial)^m\psi$ that one obtains 
$\partial_n S = {\cal B}_n(P) = \lambda^n_{+}$ where the subscript (+) 
refers now to powers of $P$ (note $\epsilon\partial_n\psi/\psi \to 
\partial_n S$). Thus $B_n = L^n_{+}\to {\cal B}_n(P) = \lambda^n_{+} = 
\sum_0^nb_{nm}P^m$ and the KP hierarchy goes to 
$(\bullet\diamondsuit\bullet)\,\, \partial_n P = 
\partial {\cal B}_n$
which is the dKP hierarchy
(note $\partial_n S = {\cal B}_n\Rightarrow \partial_n P = 
\partial{\cal B}_n$).
The action $S$ in (\ref{AAN}) can be computed from (\ref{AAG}) in the 
limit $\epsilon \to 0$ as
\be
\label{action}
S = \sum_{1}^{\infty} T_n \lambda^n - \sum_{1}^{\infty} {\partial_mF 
\over m} \lambda^{-m}
\ee
where the function $F=F(T)$ (free energy) is defined by  
\be 
\label{tau}
\tau = \exp \left[ {1 \over \epsilon^2} F(T) \right] \ee The formula 
(\ref{action}) then solves the dKP hierarchy 
$(\bullet\diamondsuit\bullet)$, i.e. $P={\cal B}_1 = 
\partial S$ and
\be
\label{B}
{\cal B}_n = \partial_n S =
\lambda^n - \sum_{1}^{\infty} {F_{nm} \over m} \lambda^{-m}  \ee 
where $F_{nm} = \partial_n\partial_m F$ which play an important role in 
the theory of dKP.
\\[3mm]\indent
Now following \cite{ta} one writes the differential Fay identity (\ref{AAL}) 
with $\epsilon\partial_n$ replacing $\partial_n$, looks at logarithms, 
and passes $\epsilon\to 0$ (using (\ref{tau})).  Then
only the second order derivatives survive, and one 
gets the dispersionless differential Fay identity 
\be \sum_{m,n=1}^{\infty}\mu^{-m}\lambda^{-n}\frac{F_{mn}}
{mn} = \log \left(1- \sum_1^{\infty}\frac{\mu^{-n}-\lambda^{-n}}{\mu-\lambda} 
\frac{F_{1n}}{n} \right)  
\label{AAT}
\ee
Although (\ref{AAT}) only uses a subset of the Pl\"ucker relations defining 
the KP hierarchy it was shown in \cite{ta} that this subset is sufficient to 
determine KP; hence (\ref{AAT}) characterizes the function $F$ for dKP. 
Following \cite{ch,cl}, we now derive a dispersionless limit of the Hirota 
bilinear equations
$({\bf HE})$, which we call the dispersionless 
Hirota equations. We first note from (\ref{action}) and $(\bullet
\heartsuit\bullet)$
that $F_{1n} = nP_{n+1}$
so
\be
\sum_1^{\infty}\lambda^{-n}\frac{F_{1n}}{n} = \sum_1^{\infty}P_{n+1} 
\lambda^{-n} = \lambda - P(\lambda) 
\label{AAU}
\ee
Consequently the right side of (\ref{AAT}) becomes $\log[\frac{P(\mu) - 
P(\lambda)}{\mu-\lambda}]$ and for $\mu\to \lambda$ with $\dot{P} = 
\partial_{\lambda}P$ we have
\be
\log\dot{P}(\lambda) = \sum_{m,n=1}^{\infty}\lambda^{-m-n}\frac{F_{mn}} 
{mn} = \sum_{j=1}^{\infty} \left(\sum_{n+m=j} {F_{mn} \over mn} \right) 
\lambda^{-j} 
\label{AAV}
\ee
Then using the elementary Schur polynomial defined in (\ref{AAH}) and 
$(\bullet\heartsuit\bullet)$, we obtain
\be
\dot{P}(\lambda) = \sum_0^{\infty} \chi_j(Z_2, \cdots,Z_j) \lambda^{-j} 
= 1 + \sum_1^{\infty}F_{1j} 
\lambda^{-j-1};\,\,
Z_i = \sum_{m+n=i} {F_{mn} \over mn}\,\,\,\,(Z_1 = 0) 
\label{AAW}
\ee 
Thus we obtain the dispersionless Hirota 
equations, 
\be \label{F}
F_{1j} = \chi_{j+1}(Z_1=0,Z_2, \cdots,Z_{j+1})  
\ee 
These can 
be also derived directly from 
the Hirota equations $({\bf HE})$ with (\ref{tau}) in the limit 
$\epsilon \to 0$ or by expanding (\ref{AAV}) in powers of $\lambda^{-n}$
as in  \cite{ch,cl}).  
The equations (\ref{F}) then characterize dKP.
\\[3mm]\indent
It is also interesting to note that the dispersionless Hirota equations 
(\ref{F})  can be regarded as algebraic equations for 
``symbols" $F_{mn}$, which are defined via (\ref{B}), i.e. \be
{\cal B}_n := \lambda^n_+= \lambda^n - \sum_1^{\infty}\frac{F_{nm}}{m} 
\lambda^{-m}
\label{AAY}
\ee
and in fact
\be
F_{nm} = F_{mn} = Res_P[\lambda^m d \lambda^n_+]  
\label{AAZ} \ee 
Thus 
for $\lambda,\,\,P$ given algebraically as in $(\bullet\heartsuit\bullet)$,
with no a priori connection to dKP, and for ${\cal B}_n$ defined as in
(\ref{AAY}) via a formal collection of symbols with two
indices $F_{mn}$, it follows that the dispersionless Hirota equations 
(\ref{F}) are nothing but polynomial identities among 
$F_{mn}$.  In particular one has from \cite{ch}
\begin{itemize}
\item
(\ref{AAZ}) with (\ref{F}) completely characterizes and solves the
dKP hierarchy.
\end{itemize}
\indent
Now one very natural way of developing dKP begins with 
$(\bullet\heartsuit\bullet)$ and 
$(\bullet\diamondsuit\bullet)$ since
eventually the $P_{j+1}$ can serve as universal coordinates (cf. here 
\cite{ae} for a discussion of this in connection with topological field 
theory = TFT). This point of view is also natural in terms of developing 
a Hamilton-Jacobi theory involving ideas from the hodograph $-$ Riemann 
invariant approach (cf. \cite{ci,gh,kn,kq,kr} and in 
connecting
NKdV ideas to TFT, strings, and quantum gravity.
It is natural here to work with $Q_n = (1/n){\cal B}_n$ and note that 
$\partial_n S = {\cal B}_n$ corresponds to $\partial_n P = \partial 
{\cal B}_n = n\partial Q_n$.
In this connection one often uses different time variables, say $T'_n = 
nT_n$, so that $\partial'_nP = \partial Q_n$, and $G_{mn} = F_{mn}/mn$ is 
used in place of $F_{mn}$. Here however we will retain the $T_n$ notation 
with $\partial_n S = nQ_n$ and $\partial_n P = n\partial Q_n$ since one 
will be connecting a number of formulas to standard KP notation. Now given 
$(\bullet\heartsuit\bullet)$ and $(\bullet\diamondsuit\bullet)$ the equation
$\partial_n P = n\partial Q_n$ corresponds to Benney's moment equations 
and is equivalent to a system of Hamiltonian equations defining the dKP 
hierarchy (cf. \cite{ci,kn});
the Hamilton-Jacobi equations are 
$\partial_n S = nQ_n$ with Hamiltonians $nQ_n(X, P=\partial S)$). 
There is now an important formula involving the functions $Q_n$ from
\cite{kn}, namely
the generating function of $\partial_P Q_n(\lambda)$ is given by 
\be 
K(\mu,\lambda)=
\frac{1}{P(\mu) - P(\lambda)} = \sum_1^{\infty}\partial_P Q_n(\lambda) 
\mu^{-n}  
\label{ABF}
\ee
In particular one notes 
\be \oint_{\infty} {\mu^n \over 
P(\mu) - P(\lambda)} d\mu = \partial_P Q_{n+1}(\lambda) \ , 
\label{canonicalT}
\ee
which gives a key formula in the Hamilton-Jacobi method for the dKP \cite{kn}. 
Also note here that the function 
$P(\lambda)$ alone provides all the information necessary for the dKP theory. 
It is proved in \cite{ch} that 
\begin{itemize}
\item
The kernel formula (\ref{ABF}) is
equivalent to the dispersionless differential Fay identity (\ref{AAT}).
\end{itemize}
\indent
The proof uses
\be \label{ShP}
\partial_P Q_n = \chi_{n-1}(Q_1,\cdots,Q_{n-1}) 
\ee 
where $\chi_n(Q_1,\cdots,Q_n)$ can be expressed as a polynomial in $Q_1 = P$
with the coefficients given by polynomials in the $P_{j+1}$.
Indeed
one shows that $\chi_n=\partial_PQ_{n+1}$ via a determinant construction
and this leads to the observation that the $F_{mn}$ can be 
expressed
as polynomials in $P_{j+1} = F_{1j}/j$.
Thus
the dispersionless Hirota
equations can be solved totally algebraically via $F_{mn} = 
\Phi_{mn}(P_2,P_3,\cdots,P_{m+n})$ where $\Phi_{mn}$ is a polynomial in 
the $P_{j+1}$ so the $F_{1n} = nP_{n+1}$ are generating elements for the 
$F_{mn}$, and serve as universal coordinates.  Indeed
formulas such as (\ref{ShP})
indicate that in fact dKP theory can be characterized 
using only elementary Schur polynomials since these provide all the 
information necessary for the kernel (\ref{ABF}) or equivalently for the 
dispersionless differential Fay identity. This amounts
also to observing that in the passage from KP to dKP only certain Schur 
polynomials survive the limiting process $\epsilon\to 0$. Such terms 
involve second derivatives of $F$ and these may be characterized in terms of 
Young diagrams with only vertical or horizontal boxes. This is also related 
to the explicit form of the hodograph transformation where one needs only 
$\partial_P Q_n = \chi_{n-1}(Q_1,\cdots,Q_{n-1})$ and the $P_{j+1}$ in the 
expansion of $P$ (cf. \cite{ch}).
Given KP and dKP theory we can now discuss nKdV or dnKdV easily although
many special aspects of nKdV for example are not visible in KP.  In
particular for the $F_{ij}$ one will have $F_{nj} = F_{jn} = 0$
for dnKdV.
We note also (cf. \cite{km}) that from (\ref{ShP}) one has
\be
\frac{1}{P(\mu)-P(\lambda)} = \sum_1^{\infty}\partial_PQ_n\mu^{-n} =
\sum_0^{\infty}\chi_n(Q)\mu^{-n}=exp(\sum_1^{\infty}Q_m\mu^{-m})
\label{ABN}
\ee

\section{CAUCHY TYPE KERNELS}
\renewcommand{\theequation}{3.\arabic{equation}}\setcounter{equation}{0}

\subsection{The background kernel}

One aim of this paper is to produce a Riemann surface analogue of the 
kernel $K(\mu,\lambda)$ of (\ref{ABF}) $\sim$ (\ref{ABN}), namely
\be
K(\mu,\lambda)=\frac{1}{P(\mu)-P(\lambda)}=\sum_1^{\infty}\partial_P
Q_n(\lambda)\mu^{-n}
\label{BC}
\ee
where
\be
\partial_PQ_{n+1}=\frac{1}{2\pi i}\oint_{\infty}\frac{\mu^nd\mu}{P(\mu)
-P(\lambda)}
\label{BD}
\ee
Here $nQ_n={\cal B}_n=\partial_nS$ so the desired analogy would naturally
arise from
\be
\partial_nd{\cal S}=d\Omega_n
\label{BA}
\ee
(cf. \cite{cz} for an extensive discussion).  On the other hand $\partial_PQ_n$
suggests using $p\sim P$ (recall $P=S_X$ and $dp=d\Omega_1$ with 
$\partial d{\cal S}=d\Omega_1$ where $\partial\sim\partial_X$ with
$X\sim T_1$).  Thus, denoting by $\gamma\in \Sigma_g$ a point on the
Riemann surface,
one has heuristically for $\Omega_n(\gamma)\sim\int^{\gamma}d\Omega_n$,
the correspondences
$\Omega_n\sim {\cal B}_n=nQ_n$ and 
\be
\frac{\partial Q_n}{\partial P}\sim\frac{1}{n}\frac{\partial\Omega_n}
{\partial p}\sim\frac{1}{n}\frac{\partial\Omega_n/\partial\gamma}{\partial p/
\partial\gamma}=\frac{1}{n}\frac{d\Omega_n}{dp}=\frac{1}{n}\frac{d\Omega_n}
{d\Omega_1}
\label{BB}
\ee
Note that in referring to $Q_n(\lambda)$ for example one is thinking of
$\lambda$ near $\infty$ and the expressions (\ref{BC}) - (\ref{BD})
involve then both $\lambda$ and $\mu$ near $\infty$; we think of 
$\lambda\sim k$ near $\infty$ and $z=\lambda^{-1}\sim k^{-1}$.  The
analogue to (\ref{BC}) is now
\be
{\cal K}(\mu,\lambda)=\sum_1^{\infty}\frac{d\Omega_j}{d\Omega_1}(\lambda)
\frac{\mu^{-j}}{j}
\label{BE}
\ee
It will be seen that (\ref{BE}) can be related to a local Cauchy kernel
on a Riemann surface expressible in terms of $d_{\zeta}log\,E(\zeta,z)$ where
$E$ is the Fay-Klein prime form, and this will be our first project.

\subsection{Theta functions and the prime form}

In \cite{cn} we organized this material in two forms, one from a physics
viewpoint following \cite{af,ag,ac} and another following \cite{fe,ma}.
In a sense \cite{fe} is surely the basic reference here but we will 
develop this material in a somewhat shortened approach following \cite{al}
(the notation of \cite{al,ag,sc} is equivalent and this gives us 
recourse to a number of formulas immediately, without worrying about
factors of $2\pi i$ etc.).  We refer generally to \cite{al,ag,bb,ca,cn,
dc,dd,fe,fz,ga,ge,gd,ha,hb,mb,ma,ob} for information on Riemann surfaces,
theta functions, half differentials, etc.  Thus begin with $\Sigma_g$ a 
compact Riemann surface of genus $g$ with a canonical homology basis
$(A_i,B_i)$ and a complex atlas $(U_{\alpha},z_{\alpha})$.  Divisors
have the form $\sum n_iP_i$ with $deg(D)=\sum n_i$ and $D\sim D'$ if
$D-D'=(f)$ is the divisor of a meromorphic function (principal divisor).
One wants to consider spinors of the form $T\sim 
f_{\alpha}(z_{\alpha})dz^{\lambda}_{\alpha}$ where $\lambda$ is an integer
or a half-integer.  For integer $\lambda,\,\,
T_1/T_2$ is a meromorphic function
so $(f^1)-(f^2)$ is a principal divisor ($T_i\sim f^i$)
and $D^{\lambda}\sim$ equivalence
class of divisors of $\lambda$ differentials.  $D^1$ is the canonical class
$K_{\Sigma}$ of one differentials
having degree $2g-2$ (cf. Section 2.1).
For $\lambda=1/2$ the transition functions $(U_{\alpha},z_{\alpha})\to
(U_{\beta},z_{\beta})$ is $g_{\alpha\beta}=(dz_{\alpha}/dz_{\beta})^{1/2}$
and the cocycle condition $g_{\alpha\beta}g_{\beta\gamma}g_{\gamma
\delta}=1$ must be satisfied.  This leads to $2^{2g}$ inequivalent divisor
classes $D_{\alpha}^{1/2}$ parametrized by a 
(characteristic) pair $\alpha=(\alpha_1,\alpha_2)$ of $g$-dimensional vectors
$\alpha_i\in\{0,\frac{1}{2}\}^g\,\,(i=1,2)$ with components $0$ or $1/2$.
We recall that ${\cal A}_i(P)=\int_{P_o}^Pd\omega_i$ where $P_0\not=
P_{\infty}$ and $d\omega_i\sim$ basis of holomorphic differentials
satisfying $(\spadesuit)\,\,\oint_{A_i}d\omega_j=\delta_{ij};\,\,
\oint_{B_i}d\omega_j=B_{ij};\,\,\Im(B_{ij})>0$.
In \cite{fe} one uses $\oint_{A_k}d\omega_j=2\pi i\delta_{jk}$
and takes $\Re(\oint_{B_k}d\omega_j<0$ (see Remark 3.1 below).  We recall
that the Jacobian $J(\Sigma_g)\sim {\bf C}/({\bf Z}^g+B{\bf Z}^g)$ and
the Abel map ${\cal A}$ extends to divisors $D=\sum n_iP_i$ via ${\cal A}
(D)=\sum n_i{\cal A}(P_i)$.
One knows that $D$ is a principal divisor if and only if ${\cal A}(D)=0$
(modulo the lattice).  Further ${\cal A}(K_{\Sigma})\sim 2\Delta$
where $\Delta\sim$ vector of Riemann constants (cf. Section 2.1 where we have
used $K\sim -\Delta$ to agree with other notations).
\\[3mm]\indent
Now the equivalence classes of $1/2$ differentials turn out to be characterized
by $D_{\alpha}^{1/2}=\Delta-B\alpha_1-\alpha_2$ and to $D_{\alpha}^{1/2}$
one associates a theta function
\be
\Theta[\alpha](z)=\sum_{n\in{\bf Z}^g}exp[i\pi (n+\alpha_1)B(n+\alpha_1)^T
+2\pi i(z+\alpha_2)(n+\alpha_1)^T]
\label{WA}
\ee
This is holomorphic on ${\bf C}^g$ and quasiperiodic via (sometimes we 
write $z$ for $z^T$)
\be
\Theta[\alpha](z+Bn+m)=exp\left(-i\pi nBn^T-2\pi n(z+\alpha_2)+2\pi im\alpha_1
\right)
\Theta[\alpha](z)
\label{WB}
\ee
Further $(\bullet)\,\,\Theta[\alpha](-z)=(-1)^{4\alpha_1\alpha_2}\Theta
[\alpha](z)$ and $\alpha$ is called even or odd depending on whether
$\Theta[\alpha](z)$ is even or odd.  We recall next by the Riemann
vanishing theorem that the zeros of $\Theta[\alpha](z)$ form the set of
$z\in {\bf C}^g$ represented via
\be
z={\cal A}\left(D_{\alpha}^{1/2}-\sum_1^{g-1}P_i\right)=\Delta-B\alpha_1
-\alpha_2-\sum_1^{g-1}{\cal A}(P_i)
\label{WC}
\ee
($P_1,\cdots,P_{g-1}$ being arbitrary).  For odd $\alpha,\,\,\Theta[\alpha](0)
=0,$ so by (\ref{WC}) there exists $g-1$ points $P_{\alpha,i}\in\Sigma_g$
such that $D_{\alpha}^{1/2}=\sum_1^{g-1}P_{\alpha,i}$.  In fact a $1/2$
differential $h_{\alpha}$ with divisor $\sum_1^{g-1}P_{\alpha,i}$ can
be obtained via
\be
h_{\alpha}(P)^2=\sum\left(\frac{\partial\Theta[\alpha]}{\partial z_j}\right)
(0)d\omega_j(P)
\label{WD}
\ee
Then the (Fay-Klein) prime form of $\Sigma_g$ is defined as the multivalued
$-1/2\times -1/2$ differential on $\Sigma_g\times \Sigma_g$ given by
\be
E(P,P')=-E(P',P)=\frac{\Theta[\alpha]({\cal A}(P-P'))}{h_{\alpha}(P)
h_{\alpha}(P')}
\label{WE}
\ee
(note ${\cal A}(P-P')={\cal A}(P)-{\cal A}(P')=(\int_{P'}^Pd\omega_j)$ and 
$\alpha\sim$ arbitrary odd characteristic).  This has the following
properties:
\begin{itemize}
\item
$E(P,P')$ has a single zero at $P=P'$
\item
If $z_1,\,z_2$ are the coordinates of $P_1,\,P_2$ then near $P_1=P_2$
$$
E(P_1,P_2)=\frac{z_1-z_2}{\sqrt{dz_1}\sqrt{dz_2}}(1+O(z_1-z_2)^2))$$
\item
As a function of $P$ the prime form is single valued around the $A$ cycles
but multivalued around the $B$ cycles; upon going around $B_j$ one has
$E(P,P')\to E(P,P')exp[-i\pi B_{jj}-2\pi i{\cal A}_j(P-P')]$
\end{itemize}
One can also use $E$ to 
construct a differential of third kind with first order poles
at $P_1$ and $P_2$ obtained via
\be
d\omega(P_1,P_2)=d_Plog\left(\frac{E(P,P_1)}{E(P,P_2)}\right)
\label{WH}
\ee
Further, given that $D=P_1 +\cdots + P_n-Q_1-
\cdots - Q_n$ is the divisor of a meromorphic function one can
express this function via 
\be
f(z) = \frac{\prod_1^nE(z,P_i)}{\prod_1^nE(z,Q_i)}
\label{HH}
\ee
\indent {\bf REMARK 3.1}$\,\,$
The notation in \cite{fe} is slightly different in that $\theta(x)$ is
defined via (we use small $\theta$ and small $b$
for \cite{fe})
\be
\theta(z')=\theta[0](z')=\sum exp\left(\frac{1}{2}mbm^T+z'm^T\right);\,\,
\theta{\alpha \brack \beta}(z')=
\label{LB}
\ee
$$=\sum e^{\frac{1}{2}(m+\alpha)b(m+\alpha)^T+(z'+2\pi i\beta)(m+\alpha)^T}
=e^{\frac{1}{2}\alpha b\alpha^T+(z'+2\pi i\beta)\alpha^T}\theta(z'+e')$$
where $e'=2\pi i\beta+\alpha b=(\beta,\alpha){2\pi i \choose b}$ and the
period matrix $b$ is symmetric with $\Re b<0$.  In this context
\be
E'(x',y')=\frac{\theta[\delta](y'-x')}{h'_{\delta}(x')h'_{\delta}(y')};\,\,
(h'_{\delta})^2(x')=\sum_1^g\left(\frac{\partial\theta[\delta](0)}{\partial
z'_i}\right)d\omega'(x')
\label{LA}
\ee
Comparing with the theta function of (\ref{WA}) etc. we see that
$b\sim 2\pi iB$ and $z'\sim 2\pi iz$ with $\alpha\sim\alpha_1$ and $\beta\sim
\alpha_2$.  We will want to use later a formula from \cite{fe} involving 
$d_{x'}log\,E'(x',y')$ as a local Cauchy kernel and since the formulas
connected with this have an invariant meaning on $\Sigma_g$ we will simply
transplant the notation to $E$ and $\Theta$ as in (\ref{WA}) and (\ref{WE})
(note e.g. $z'=\alpha z\Rightarrow f_zdz=f_{z'}dz'$).
A tedious calculation could of course be made using the comparisons 
just indicated but that seems unnecessary.
\\[3mm]\indent
Thus keeping our standard notation with $\Im B>0$ one notes from \cite{fe}
that
${\cal Z}_p(x) = (d/dx)log\,E(x,p)$
has a pole of residue $1$ at $x=p$ and $[{\cal Z}_b(x) -
{\cal Z}_a(x)]dx = d\omega_{b-a}(x) = \int^b_a\
d_xd_ylog\,E(x,y)dy$ (cf. (\ref{BH})).
It is shown in \cite{fe} that for $f$ holomorphic
in a neighborhood $U$ of $p$ one has
\be
f(p) = \frac{1}{2\pi i}\int_{\partial U}f(x){\cal Z}_p(x)dx
\label{LK}
\ee
so $Z_p$ is a local Cauchy kernel.
Next recall for $s\geq 1,\,\,d\Omega_s=(-sz^{-s-1}-\sum_1^{\infty}
q_{ms}z^{m-1})dz$ with $q_{ms}=q_{sm}$ (via the Riemann bilinear
relations) or alternatively, $d\Omega_s=(s\lambda^{s-1}+\sum_1^{\infty}
q_{ms}\lambda^{-m-1})d\lambda\,\,(\lambda\sim k)$.  Similarly
$d\omega_j=-\sum_1^{\infty}\sigma_{jm}z^{m-1}dz=\sum_1^{\infty}
\sigma_{jm}\lambda^{-m-1}d\lambda$.  Consider now
$\hat{\omega}_2(z,\zeta) = \partial_{\zeta}
\partial_z log E(z,\zeta)
= \partial_{\zeta}(E_z/E) = (E_{z\zeta}/E) - (E_z
E_{\zeta}/E^2)$ and via $E_z\not= 0,\,\,E_{\zeta}\not= 0$ at $z=\zeta$
the last term looks like $1/(z-\zeta)^2$ (note also $\hat{\omega}_2
dz$ has zero $A_i$ periods - cf. \cite{ma}).  This is a standard way
of generating a differential with a second order pole. 
From this one picks up
differentials of the second kind with poles of order $n+1$ at $z=\zeta$
via $\hat{\omega}_{n+1} = \partial^{n-1}_{\zeta}\hat{\omega}_2(z,\zeta)/
n!\,\,(n=2,\cdots)$ for example and we note that (cf. \cite{ag,sc})
\be
\oint_{B_i}\hat{\omega}_{n+1}(z,\zeta)dz = \frac{2\pi i}{n!}D_{\zeta}^
{n-1}f_i(\zeta)
\label{HJ}
\ee
where $d\omega_i = f_id\zeta$.  Further (cf. (\ref{WH}) with $z\sim P$)
\be
\oint_{\vec{B}}\,\,d\omega(z,P,\hat{P}) = 2\pi i\int_{\hat{P}}^Pd\omega
\label{HK}
\ee
where $\vec{B}\sim (B_1,\cdots,B_g)$ and $d\omega\sim (d\omega_1,\cdots,
d\omega_g)$ represents the standard holomorphic differentials.  Here the
relation (\ref{HK}) follows from standard bilinear identities as does
\be
\oint_{B_k}\hat{\omega}_2dz = 2\pi i\frac{d\omega_k}{d\zeta}
\label{HL}
\ee
and (\ref{HJ}) follows from (\ref{HL}). 
\\[3mm]\indent
Now one can write ($z\sim 1/k$)
\be
\hat{\omega}_2(z,\zeta)dzd{\zeta}
= d_zd_{\zeta}\,log\,E(z,\zeta) = -\sum_1^{\infty}
d\Omega_s(z)\zeta^{s-1}d\zeta=
\label{HN}
\ee
$$=-\sum_1^{\infty}\left[-sz^{-s-1}-\sum_1^{\infty}q_{ms}z^{m-1}\right]
\zeta^{s-1}d\zeta=\sum_1^{\infty}sz^{-s-1}\zeta^{s-1}dzd\zeta+$$
$$+\sum_{m,s=1}^{\infty}z^{m-1}\zeta^{s-1}dzd\zeta
=\frac{dzd\zeta}{(\zeta-z)^2}+\sum_{m,s=1}^{\infty}q_{ms}
z^{m-1}\zeta^{s-1}dzd\zeta$$
\be
\hat{\omega}_{n+1}(z,\zeta)dzd\zeta =
\partial_{\zeta}^{n-1}\frac{\hat{\omega}_2(z,\zeta)}{n!}dzd\zeta =
\label{HHN}
\ee
$$= -[\frac
{(n-1)!}{n!}d\Omega_n(z) +\sum_{n+1}^{\infty}{s-1 \choose n}d\Omega_s(z)
\zeta^{s-n}]d\zeta
$$
This leads to
\be
d\Omega_n(z)d\zeta = -n\hat{\omega}_{n+1}(z,0)dzd\zeta = -\frac{1}{(n-1)!}
\partial_{\zeta}^{n-1}\hat{\omega}_2(x,\zeta)|_{\zeta=0}dzd\zeta\,\,(n\geq 1)
\label{HNN}
\ee
\be
\Omega_{sm}\sim\oint_{B_m}d\Omega_s\sim\oint_{B_m}\hat{\omega}_{s+1}dz|_
{\zeta=0} =
\label{HO}
\ee
$$ = \frac{1}{s!}D_{\zeta}^{s-1}\oint_{b_m}\hat{\omega}_2dz|_{\zeta=0} = 
\frac{2\pi i}{s!}
D_{\zeta}^{s-1}f_m(\zeta)|_{\zeta=0}$$
\\[3mm]\indent
Now to apply this to the kernel ${\cal K}$ of (\ref{BE}) we write
$\zeta\sim (1/\lambda)$ and $z\sim (1/\mu)$ with $\lambda,\,\mu\to
\infty$.  One can write ${\cal K}$ in the form
\be
{\cal K}(\mu,\lambda)=\sum_1^{\infty}\frac{d\Omega_j}{d\Omega_1}(\lambda)
\frac{\mu^{-j}}{j}\equiv \sum_1^{\infty}\frac{d\Omega_j}{d\Omega_1}
(\zeta)\frac{z^j}{j}=\tilde{{\cal K}}(z,\zeta)
\label{BG}
\ee
We can also write from above
\be
{\cal Z}_z(\zeta)=d_{\zeta}log\,E(\zeta,z);\,\,[{\cal Z}_z(\zeta)-
{\cal Z}_w(\zeta)]d\zeta=
\label{BH}
\ee
$$=d\omega_{z-w}=\left(\int_w^z\partial_{\zeta}\partial_ylogE
(\zeta,y)dy\right)d\zeta$$
From (\ref{HN}) we have
\be
\left(\int_w^z\partial_{\zeta}\partial_ylog\,E(\zeta,y)dy\right)d\zeta=
\left[\int_w^z\left(\frac{dy}{(y-\zeta)^2}+\sum q_{ms}\zeta^{m-1}
y^{s-1}\right)dy\right]d\zeta=
\label{BI}
\ee
$$=\left[\frac{1}{w-\zeta}-\frac{1}{z-\zeta}+
\sum_{m,s=1}^{\infty}\frac{q_{ms}}{s}
\zeta^{m-1}(z^s-w^s)\right]d\zeta$$
This implies
\be
{\cal Z}_z(\zeta)=\left[\frac{1}{\zeta-z}+\sum_{m,s=1}^{\infty}
\frac{q_{ms}}{s}\zeta^{m-1}z^s\right]d\zeta
\label{BJ}
\ee
Now write out $\tilde{{\cal K}}(z,\zeta)$ as
\be
\tilde{{\cal K}}(z,\zeta)=\frac{1}{d\Omega_1(\zeta)}\sum_1^{\infty}
\frac{z^j}{j}\left[-j\zeta^{-j-1}-\sum_1^{\infty}q_{mj}\zeta^{m-1}\right]
d\zeta=
\label{BK}
\ee
$$=\frac{1}{d\Omega_1(\zeta)}\sum\left[\frac{1}{\zeta}-\frac{1}
{\zeta-z}-\sum_{m,j=1}^{\infty}\frac{q_{mj}}{j}\zeta^{m-1}z^j\right]d\zeta=
\frac{1}{d\Omega_1(\zeta)}\left[\frac{d\zeta}{\zeta}-
{\cal Z}_z(\zeta)\right]$$
(note $(1/\zeta)-(1/(\zeta-z))=-\sum_1^{\infty}(z^j/\zeta^{j+1})$).
This leads to
\\[3mm]\indent {\bf THEOREM 3.2.}$\,\,$ The kernel ${\cal K}$ of (\ref{BE})
or equivalently $\tilde{{\cal K}}$ of (\ref{BG}) can be written in the form
(\ref{BK}) in terms of the local Cauchy kernel ${\cal Z}_z(\zeta)$.
\\[3mm]\indent {\bf REMARK 3.3.}$\,\,$  From (\ref{BC}) we have
$K(\mu,\lambda)=1/(P(\mu)-P(\lambda))$ where $P\sim d\Omega_1$ on the
Riemann surface.  Then formally and heuristically on the Riemann surface
we are looking at
\be
\sum d\Omega_j(\lambda)\frac{\mu^{-j}}{j}\sim \frac{d\Omega_1(\lambda)}
{\Omega_1(\mu)-\Omega_1(\lambda)}=\frac{dp(\lambda)}{p(\mu)-p(\lambda)}
\label{BL}
\ee
In terms of $z,\,\zeta$ this leads to
\be
\sum d\Omega_j(\zeta)\frac{z^j}{j}=\frac{dp(\zeta)}{p(z)-p(\zeta)}=
\frac{d\zeta}{\zeta}-{\cal Z}_z(\zeta)
\label{BM}
\ee
Note here for consistency $\Upsilon=(d\zeta/\zeta)-(d\zeta/(\zeta-z))=
-(zd\zeta/\zeta(\zeta-z))$ and for $\zeta=1/\lambda,\,\,z=1/\mu,$ and
$d\zeta=-(1/\lambda^2)d\lambda$ we obtain $\Upsilon=d\lambda/(\mu-\lambda)$
as expected from (\ref{BC}).

\section{KERNELS BASED ON $\psi\psi^*$}
\renewcommand{\theequation}{4.\arabic{equation}}\setcounter{equation}{0}

We go back to the BA function for KP as in (\ref{psi}) so that
$\psi(\vec{t},k)$ is meromorphic in $\Sigma_g/\infty$ with simple
poles at $D\sim P_1,\cdots,P_g$ and no other singularities in $\Sigma_g/\infty$
(here $\vec{t} = (t_i)$ with $t_1=x,\,\,t_2=y,\,\,t_3=t,$ etc.).  We use
$k^{-1}$ as the local coordinate at $\infty$ so $\psi\sim exp[\xi
(\vec{t},k)]\cdot (1+\sum_1^{\infty}\chi_i k^{-i})$ for $|k|$ large
where $\xi(\vec{t},k) = \sum_1^{\infty}t_ik^i
\sim q(k)$.  For $t_{2i} = 0$
this is a KdV situation.
The BA conjugate differential $\psi^{\dagger}(\vec{t},\mu)$ 
can be defined via $(\clubsuit)\,\,\psi^{\dagger}=\psi^*d\hat{\Omega}
\sim exp[-\xi(\vec{t},k)](1+\sum_1^{\infty}\xi_i^*k^{-i})
(1+(\beta/k^2)+\cdots)dk$ 
where
$d\hat{\Omega}$ is the 
meromorphic differential indicated in Section 2 with zeros at $D+D^*$
and a double pole at $\infty$.
Then one has
\be
\oint_C\psi(k,\vec{t})\psi^{\dagger}(k,\vec{t'})dk =
\oint_C\psi(k,\vec{t})\psi^*(k,\vec{t'})d\hat{\Omega}= 0
\label{GH}
\ee
for $C$ a small contour around $\infty$ (Hirota bilinear identity) and
\be
\int_{-\infty}^{\infty}\psi(k,\vec{t})\psi^{\dagger}(k',\vec{t})dx
=2\pi i\delta(k-k')
\label{GI}
\ee
for $\Im p(k) = \Im p(k')$.  In (\ref{GI}) the contour could apparently
be a closed curve through $\infty$ for example (which would become a straight
line in the - degenerate - scattering situation).  This point is however
consistently overlooked and should be further examined (cf. also
\cite{cc,ct}).  
Next the Cauchy-Baker-Akhiezer
(CBA) kernel $\omega(k,k',\vec{t})$ is defined via
(we give an expanded form later and cf. also \cite{ra,za} for kernels)
\be
d\omega(k,k',x,y,t,\cdots) = \frac{1}{2\pi i}\int_{\pm\infty}^x
\psi(k,x',y,t,\cdots)\psi^{\dagger}(k',x',y,t,\cdots)dx'
\label{GJ}
\ee
According to \cite{ge,gd,gf}
this kernel is to have the following 
properties:  ({\bf A})  $d\omega(k,k',\vec{t})$
is a function in $k$ and a one form in $k'$  ({\bf B})  $d\omega(k,k',
\vec{t})$ is meromorphic in $k$ in $\Sigma_g/\infty$ with simple poles
at $P_1,\cdots,P_g,k'$ ({\bf C})  As a function of $k',\,\,d\omega$ is
meromorphic in $\Sigma/\infty$ with one pole $k$ and zeros at $P_1,\cdots,
P_g$ ({\bf D})  $d\omega(k,k',\vec{t}) = O(exp[\xi(k,\vec{t})])$ as $k\to
\infty$ ({\bf E})  $\omega(k,k',\vec{t}) = O(exp[-\xi(k',\vec{t})])$ as
$k'\to\infty$ ({\bf F})  $\omega(k,k',\vec{t})\sim(dk'/2\pi i(k'-k))$ as
$k\to k'$.  
\\[3mm]\indent
The material on the prime form is well discussed in \cite{fe,ma} 
for example (cf. also \cite{hb}) 
but the proofs in \cite{ge,gd,gf} regarding (\ref{GI}), (\ref{GJ}),
and some of the properties ({\bf A}) - ({\bf F}) 
of the CBA kernel are somewhat
unclear so we will give a little discussion in various contexts
(cf. also \cite{ya}).
Formula (\ref{GH}) with $\psi^*$ in place of $\psi^{\dagger}$ (Hirota
bilinear identity) can be proved in various traditional manners (cf.
\cite{ca,cc}) so we omit comment.  Now look at the integrands in
(\ref{GI}) and (\ref{GJ}) for large $k,k'$ and $\Im p(k) = \Im p(k')$,
namely, $\psi(k,\vec{t})\psi^{\dagger}(k',\vec{t})\sim exp[\sum
t_n(k^n-k'^n)](1+\sum\xi_ik^{-i})(1+\sum\xi_j^*k'^{-j})
(1+(\beta/k)+\cdots)dk'$.
As $x = t_1$ varies one has a multiplier $exp[x(k-k')]$ with 
$|exp[x(k-k')]| = exp[x\Re(k-k')]$ (note $dp\sim -idk$ so for large
$k,\,k', \Im p(k') = \Im p(k)\sim \Re k = \Re k'$).  Thus $exp[x
(k-k')]\sim exp[ix(\Im k - \Im k')]$ and (\ref{GI}) has a Fourier
intergal flavor.  On the other hand from $\partial_n\psi = B_n\psi$ and
$\partial_n\psi^{\dagger} = -B_n^*\psi^{\dagger}$ 
one obtains 
($\Im p(k') = \Im p(k)$)
\be
\partial_n\int_{-\infty}^{\infty}\psi\psi^{\dagger}dx = \int_{-\infty}^{\infty}
[(B_n\psi)\psi^{\dagger} - \psi(B_n^*\psi^{\dagger})]dx = 0
\label{GP}
\ee
provided integration by parts is permitted (not perhaps obviously valid
but one can assume it under reasonable circumstances).
Note that no $x,y,t$ dependence is introduced in passing from 
$\psi^*$ to $\psi^{\dagger}$ so $\partial_n\psi^{\dagger}=-B_n^*\psi^{\dagger}$
follows from $\partial_n\psi^*=-B_n^*\psi^*$.
Further for large $t_n$ and $k'\not= k,\,\,|\psi\psi^*|\sim O(|exp
[t_n(k^n-k'^n)]|) = exp[t_n\Re(k^n-k'^n)]\to 0$ for some $n$ when 
$t_n\to \infty$ or $-\infty$ (with $\Im p(k') = \Im p(k)$ or not).  This
(with (\ref{GP})) implies that the integral in (\ref{GI}) is $0$ for
$k'\not= k$ and $\Im p(k') = \Im p(k)$.  
To show that (\ref{GI}) actually gives a delta function
for $\Im p(k') = \Im p(k)$ one is referred 
in \cite{ge} to \cite{kb} for a discrete
version whose proof is said to be extendable (the result is
of course natural,
up to normalization).
Finally it is clear that
the $\pm \infty$ limit of the integral in (\ref{GJ}) may change when
$k$ crosses the path $\Im p(k) = \Im p(k')$.  In this regard recall
$|exp[x(k'-k)]| = exp[x(\Re k' - \Re k)]$ and $\Im p(k)\sim -\Re k$ for
$k$ large.  Hence $\omega$ in (\ref{GJ}) can have a jump discontinuity
across the curve $C:\,\,\Im p(k) = \Im p(k')$.  To see this and to show
that $\omega(k,k')$ is nevertheless continuous for $k\not= k'$ let $I
\subset C$ be a small arc seqment and $D\supset I$ a small open set with
boundary $\partial D$.  Then $\int_D\bar{\partial}\omega dkd\bar{k}
=\int_{\partial D}\omega dk$ by Stokes theorem.  Think of $I$ as a small
straight line segment with $D$ shrunken down around $I$ to be lines
above and below with little end curves.  One obtains in an obvious notation
\be
\int_D\bar{\partial}d\omega dkd\bar{k} = \int_I(d\omega_{+} - d\omega_{-})dk =
\label{GQ}
\ee
$$=\frac{1}{2\pi}\int_I\int_{-\infty}^{\infty}\psi(k)\psi^{\dagger}(k')dxdk = 
\left\{
\begin{array}{cc}
0 & for\,\,k'\not\in I\\
1 & for \,\,k'\in I
\end{array}
\right.
$$
via (\ref{GI}) and (\ref{GJ}) (using the change in integration limit in
(\ref{GJ})).  The other properties in ({\bf A}) - ({\bf E}) are more or
less natural.  
Property ({\bf F})  is suggested by
e.g. $(1/2\pi i)\int_{-\infty}^x\psi\psi^{\dagger}dx'dk'\sim
[1/2\pi i(k-k')]\cdot\int^x_{-\infty}exp[x'(k-k')]\cdot O(1)dx'dk'$  when
say $\Re k > \Re k',\,\,k'\to k$ and $x>0$ with the
expectation here that $1/(k-k')$ will emerge from
the integration (cf. Theorem 4.3).
We will indicate other formulas for $d\omega$ and $\psi^{\dagger}$ below
which accord with (\ref{GH}), (\ref{GI}), and 
({\bf A}) - ({\bf F}).
\\[3mm]\indent  Now
consider from (\ref{GJ}) $\partial_xd\omega(k,k',\vec{t}) = (1/2\pi i)
\psi(k,\vec{t})\psi^{\dagger}(k',\vec{t})$ where $\psi^{\dagger}$ is given
via $\psi^{\dagger}=\psi^*d\hat{\Omega}$ as indicated above.
Let us 
rephrase the construction of BA functions etc. now in the notation
of \cite{al} (cf. also \cite{cn,ks}).
In (\ref{psi}) for example we pick $d\Omega^1\sim
d\Omega_1 = dk + \cdots,\,\,d\Omega^j\sim d\Omega_j = d(k^j) + \cdots$ with
$\oint_{A_i}d\Omega_j = 0$ (or sometimes $\Re\oint_{A_i}d\Omega_j = 0 =
\Re\oint_{B_i}d\Omega_j$), and recall $dp\sim d\Omega_1$.
The flow variables arise via $q(k)$ in 
Section 2, but the Riemann surface contributes via the argument $xU+
yV+yW+\cdots\sim\sum t_j(\Omega_{jk})$,
where $\Omega_{jk}=\oint_{B_k}d\Omega_j,$ 
in the theta function, to establish linear flows on the Jacobian
$J(\Sigma_g)$ (note from (\ref{HNN}) that our $d\Omega_n\sim\omega_n$ in
\cite{al}).
For background here we note 
also that the zeros of $\Theta[0](z)$ are the points
$z=\Delta-{\cal A}(\sum_1^{g-1}P_i)$ with arbitrary $P_1,\cdots,P_{g-1}$.
Then given a positive divisor $D=P_1+\cdots+P_g$ the multivalued function of
$P\in\Sigma_g$ defined by $f(P)=\Theta({\cal A}(P)+\Delta-{\cal A}(D))$ 
vanishes at $P=P_1,\cdots,P_g$.  Note that for divisors satisfying
a relation $(\bullet\bullet)\,\,{\cal A}(D)=2\Delta-{\cal A}(Q_1+\cdots+
Q_{g-2})$ with arbitrary $Q_i$, we have $f(P)=\Theta(\Delta-{\cal A}
(Q_1+\cdots+Q_{g-2}+P)\equiv 0$.  Such divisors satisfying $(\bullet
\bullet)$ are called special divisors.
\\[3mm]\indent
We can write the BA function now
in the form
\be
\psi = exp(\int^P_{P_0}\sum t_nd\Omega_n)
\cdot\frac{\Theta({\cal A}(P) + \sum (t_j/2\pi i)(\Omega_{jk})+z_0)}
{\theta({\cal A}(P)+z_0)}
\label{HW}
\ee
where $z_0=-K-{\cal A}(D)=\Delta-{\cal A}(D)\sim e(D)$ in \cite{al}
and $\Theta(z)\sim\Theta[0](z)$; the only change from (\ref{psi}) is a 
factor of $1/2\pi i$ in front of the $(\Omega_{jk})$
(cf. \cite{cz} where this is also done in adjusting an argument
of \cite{ne}).
We recall as before in Section 2.1 that the path of integration $\int_{P_0}^P$
is to be the same for all factors in (\ref{HW}) and then $\psi$ is a single
valued function of $P\in \Sigma_g$.  For the dual BA function one uses a
dual divisor $D^*$ as before in Section 2.1 with $D+D^*-2Q\sim K_{\Sigma}$
where $Q\sim P_{\infty}$.  Then one can write $\psi^*$ as in (\ref{star})
with $\vec{U}$ replaced by $\sum (t_j/2\pi i)(\Omega_{jk})$, i.e.
\be
\psi^*\sim e^{-\int_{P_o}^P\sum t_nd\Omega_n}\cdot \frac{\Theta
({\cal A}(P)-\sum (t_j/2\pi i)(\Omega_{jk})+z_0^*)}{\Theta({\cal A}(P)
+z_0^*)}
\label{WI}
\ee
The differential $d\hat{\Omega}$ can be written as in \cite{al}, namely
\be
d\hat{\Omega}(P')=\frac{\Theta({\cal A}(P')+z_0)\Theta({\cal A}(P')+z_0^*)}
{E(P,P_{\infty})^2}
\label{WJ}
\ee
This is shown to be single valued in \cite{al} and replaces here an earlier
(multivalued) version of \cite{cn}.  Note also that the $1/E^2$ term can
be computed as in \cite{al} and gives rise to a $dk'$ in the numerator
as needed (see Remark 4.2).
Then (\ref{GJ}) can be written as
\be
d\omega(P,P',x,y,t,\cdots) = \frac{1}{2\pi i}\int_{\pm\infty}^x
\psi(k,x',y,t,\cdots)\psi^*(k',x',y,t,\cdots)d\hat{\Omega}(k')dx'
\label{kkernel}
\ee
and simplified as
\be
d\omega_x(P,P',\vec{t}\,) =
\frac{e^{(\int_{P_0}^P-\int_{P_0}^{P'})
d\Omega}}{2\pi i}
\cdot
\label{Ker}
\ee
$$\cdot
\frac{\Theta({\cal A}(P)+\sum (t_j/2\pi i)(\Omega_{jk})+z_0)\Theta
({\cal A}(P')-\sum (t_j/2\pi i)(\Omega_{jk})+z_0^*)\Theta({\cal A}(P')+z_0)}
{\theta({\cal A}(P)+z_0)E(P',P_{\infty})^2}$$
Hence we have
\\[3mm]\indent {\bf THEOREM 4.1}.$\,\,$  A path independent expression
for $d\omega_x$ having the essential poles and singularities 
stipulated in ({\bf A}) - ({\bf F}) can be
written as in (\ref{Ker}).
\\[3mm]
\indent {\bf REMARK 4.2}.$\,\,$
A definition of $d\omega$ via $\psi\psi^*$ (instead of $\psi\psi^{\dagger}$)
is appropriate in dealing with dispersionless limits 
but in order to have a Cauchy kernel the poles from
$\psi\psi^*$ should be eliminated (hence $\psi^{\dagger}$).  
This leads to $\psi^{\dagger}$ as in $(\clubsuit)$ and $d\omega_x$ as
in (\ref{Ker}).  The pole in $d\omega$ at $k=k'$ will emerge from the 
integration as indicated before 
(cf. also Theorem 4.3) and we will have a discontinuous Cauchy
kernel analogue in the spirit of \cite{za} (cf. also \cite{cn}).
The expression for $1/E^2$ in \cite{al} is given as
\be
\frac{1}{E(k^{-1},P_{\infty})^2}=-exp\left(\sum_{m,n=1}^{\infty}C_{nm}
\frac{k^{-m-n}}{mn}\right)dk
\label{KKK}
\ee
where
\be
C_{mn}=-\frac{1}{(n-1)!(m-1)!}\partial^n_z\partial^m_{z'}log\left.
\left(\frac{E(z,z'}{z-z'}\right)\right|_{z=z'=0}
\label{KKL}
\ee
Let us also note here, using (\ref{BI}) that
\be
C_{mn}=\frac{1}{(m-1)!(n-1)!}\partial_y^{m-1}\partial_{\zeta}^{n-1}\left.
\left[\frac{1}{(\zeta-y)^2}-\partial_{\zeta}\partial_y\,log\,E(\zeta,y)\right]
\right|_{y=\zeta=0}=
\label{KKM}
\ee
$$=\frac{-1}{(m-1)!(n-1)!}\partial_y^{m-1}\partial_{\zeta}^{n-1}\left.
\left[\sum_{p,x=1}^{\infty}q_{ps}\zeta^{p-1}y^{s-1}\right]\right|_
{y=\zeta=0}$$
leading to $C_{mn}=-q_{mn}$ for $m,n\geq 1$.
\indent
We recall now from \cite{ab} that the differential Fay identity leads to
\be
\psi^*(\vec{t},\lambda)\psi(\vec{t},\mu) =\frac{1}{\mu-\lambda}
\partial\{\frac{X(\vec{t},\lambda,\mu)\tau(\vec{t}\,)}{\tau(\vec{t}\,)}\}=
\label{KH}
\ee
$$= \frac{1}{\mu-\lambda}\partial\{e^{\sum t_j(\mu^j-\lambda^j)}
\frac{\tau(\vec{t}+[\lambda^{-1}]-[\mu^{-1}])}{\tau(\vec{t}\,)}\}$$
Given the equivalence of the dispersionless differential Fay identity
with the kernel expansion (\ref{ABF}) or (\ref{BC}) one expects
(\ref{KH}) to be at least formally useful in studying $d\omega$ in (\ref{GJ})
since (\ref{KH}) implies 
\be
d\omega\sim\frac{1}{2\pi i}\partial_{x'}^{-1}[\psi^*(x',t_n,\lambda)\psi
(x',t_n,\mu)]d\hat{\Omega}(\lambda)\sim
\label{KI}
\ee
$$\sim\frac{1}{2\pi i}\frac{1}{\mu-\lambda}\frac
{X(\vec{t},\lambda,\mu)\tau(\vec{t}\,)}{\tau(\vec{t}\,)}
d\hat{\Omega}(\lambda)$$
modulo integration or normalization factors (note (\ref{KI}) implies
$d\omega\sim d\lambda/2\pi i(\mu-\lambda)$ as $\mu\to \lambda$).   Hence one
should be able to determine easily a dispersionless limit for $d\omega$ 
(thinking of asymptotic expansions around $\infty$).  Thus
as in \cite{ch,ta} we
express $\tau$ via $\tau\sim exp(F(T)/\epsilon^2)$ and
write
\be
d\omega_{\epsilon}\sim\frac{1}{2\pi i}\frac{d\hat{\Omega}(\lambda)}
{\mu-\lambda}e^{\frac{1}
{\epsilon}\sum T_i(\mu^i-\lambda^i)}\cdot\frac{\tau(\vec{T}+
\epsilon[\lambda^{-1}]-\epsilon[\mu^{-1}])}{\tau(\vec{T}\,)}
\label{KJ}
\ee
Take logarithms to obtain then
\be
log\,d\omega_{\epsilon}\sim log\frac{1}{2\pi i}- log(\mu-\lambda) +
\frac{1}{\epsilon}\sum T_i(\mu^i-\lambda^i) +
\label{KK}
\ee
$$+ \frac{1}{\epsilon^2}\{F(\vec{T}+\epsilon [\lambda^{-1}] -\epsilon
[\mu^{-1}])-F(\vec{T}\,)\}+log\,d\hat{\Omega}$$
The next to last term can be written 
as ($\chi_n\sim$ elementary Schur functions)
\be
\frac{1}{\epsilon^2}\{e^{\sum\lambda^{-i}\frac{\epsilon\partial_i}
{i}}\cdot e^{-\sum\mu^{-1}\frac{\epsilon\partial_i}{i}}F - F\}=
\label{KL}
\ee
$$=\frac{1}{\epsilon^2}\{\sum_0^{\infty}\chi_n(\epsilon\tilde{\partial})
\lambda^{-n}\cdot\sum_0^{\infty}\chi_m(-\epsilon\tilde{\partial})\mu^{-m}F
-F\} =$$
$$=\frac{1}{\epsilon^2}\{\sum_1^{\infty}(\,\,\,)_n + \sum_1^{\infty}(\,\,\,)_m
+ \sum_1^{\infty}\sum_1^{\infty}(\,\,\,)_n(\,\,\,)_m\}$$
Now we have (cf. (cf. \cite{ch,ta})
\be
\frac{1}{\epsilon^2}\sum_1^{\infty}\sum_1^{\infty}(\,\,\,)_n(\,\,\,)_m\to
-\sum_1^{\infty}\sum_1^{\infty}\frac{F_{nm}}{nm}\lambda^{-n}\mu^{-m}
\label{KM}
\ee
and we
recall here from \cite{ch}
\be
\sum_1^{\infty}\sum_1^{\infty}\frac{F_{nm}}{nm}\lambda^{-n}\mu^{-m} =
-log(1-\frac{\mu}{\lambda})-\sum_1^{\infty}\frac{Q_n(\mu)}{\lambda^n}=
\label{KN}
\ee
$$=-log(\lambda-\mu)+log\lambda - \sum_1^{\infty}Q_n(\mu)\lambda^{-n} 
= log[\frac{P(\lambda)-P(\mu)}{\lambda-\mu}]$$
(cf. also Section 2).
Now try an expression $d\omega_{\epsilon} = 
f(\lambda,\mu)exp(R/\epsilon)d\hat{\Omega}$ which will
entrain
\be
\frac{R}{\epsilon}+log\,f\sim log(\frac{1}{2\pi i})-
log(\mu-\lambda) + 
\frac{1}{\epsilon}\sum T_i(\mu^i-\lambda^i) +
\label{KO}
\ee
$$+ \frac{1}{\epsilon^2}\{\sum_1^{\infty}(\,\,\,)_n + \sum_1^{\infty}(\,\,\,)_m
-\sum_1^{\infty}\sum_1^{\infty}(\,\,\,)_n(\,\,\,)_m\}$$
Multiply by $\epsilon$ and let $\epsilon\to 0$ to obtain
\be
R\sim\sum T_i(\mu^i-\lambda^i) + \sum_1^{\infty}\frac{F_n}{n}\lambda^{-n}
-\sum_1^{\infty}\frac{F_m}{m}\mu^{-m}
\label{KP}
\ee
which leads to 
\be
log\,f\sim log(\frac{1}{2\pi i})-log(\mu-\lambda)-log(\frac
{P(\mu)-P(\lambda)}{\mu-\lambda})
\label{KPP}
\ee
This implies via \cite{ch} and Section 2
\be
\partial_XR = (\mu-\lambda) + \sum_1^{\infty}\frac{F_{1n}}{n}(\lambda^{-n}
-\mu^{-n}) = P(\mu)-P(\lambda)
\label{KQ}
\ee
and
\be
log\,f=-log[P(\mu)-P(\lambda)]+log(\frac{1}{2\pi i})
\label{KQQ}
\ee
Thus $d\omega_{\epsilon}\sim (1/2\pi i)[1/(P(\mu)-P(\lambda))]exp\{1/
\epsilon)[S(\mu)-S(\lambda)]\}d\hat{\Omega}$
and $\partial_x\,log\,d\omega_{\epsilon}\sim P(\mu) -
P(\lambda)\,\,(\partial_x = \epsilon\partial_X)$.  
Note $\partial_XS = P$ in the notation of \cite{ch}
implies
\be
\partial_X R =  \partial_X[S(\mu)-S(\lambda)]
\Rightarrow R \sim S(\mu)-S(\lambda)
\label{KR}
\ee
One can therefore state
\\[3mm]\indent {\bf THEOREM 4.3}.$\,\,$
A dispersionless kernel analogous to $\omega$ can be modeled on 
(\ref{KI}), to be extracted from 
\be
2\pi i\,d\omega_{\epsilon}\sim\frac{d\hat{\Omega}(\lambda)}{P(\mu)-P(\lambda)}
e^{\frac{1}{\epsilon}
[S(\mu)-S(\lambda)]}
\label{KS}
\ee
(recall also $\psi\sim exp(S/\epsilon)$ and $\psi^*\sim
exp(-S/\epsilon)$ in the dispersionless theory).
Thus $2\pi i\,d\omega_{\epsilon}\sim\psi\psi^*d\hat{\Omega}$ in (\ref{KS})
gives $(\clubsuit\clubsuit)\,\,
\partial_x(log\,d\omega_{\epsilon})\sim\partial_x\frac{1}{\epsilon}
[S(\mu)-S(\lambda)]\sim P(\mu)-P(\lambda)$.
On the other hand (following \cite{cz}), given $p(\mu)=<log\psi(\vec{t},
\mu)>$ (ergodic averaging), we have from (\ref{KI}) $(\spadesuit\spadesuit)
\,\,<\partial_xlog\,d\omega>=p(\mu)-p(\lambda)$.  This indicates an
interesting relation between $K(\mu,\lambda)\sim {\cal K}(\mu.
\lambda)$ and $d\omega$ (the latter being based on $\psi\psi^{\dagger}$),
as well as a connection between e.g. $p(\lambda)$ and $P(\lambda)$.
\\[3mm]\indent {\bf REMARK 4.4.}$\,\,$  Referring now to \cite{ch} and Section
2, we recall that $(\bullet\heartsuit\bullet)$ and $\partial_nP=\partial
{\cal B}_n$ represent the dKP hierarchy, which is characterized by the
dispersionless differential Fay identity, or equivalently by the kernel
formula (\ref{ABF}).  Thus the slow variables are inserted by hand with
$\tau$ subsequently represented via (\ref{tau}), and (\ref{ABF}) arises
to characterize dKP (along with $\partial_m\partial_nF=F_{mn}$ corresponding
to $\partial_nP=\partial{\cal B}_n$).  The connection to Riemann surfaces
involves then $\partial_nd{\cal S}=d\Omega_n$ as in (\ref{BA}) with
Whitham equations based on $\partial_ndp=\partial d\Omega_n$ or $\partial_np
=\partial\Omega_n$, leading formally to $\partial^{-1}\partial_{mn}p=
\partial_m\Omega_n=\partial_n\Omega_m$ by compatibility (cf. \cite{cz}).
One can also argue via zero curvature equations as in \cite{cz,kj,ta}.
Now in \cite{al} (cf. also \cite{cz,ne}) it is shown that the
Hirota bilinear identity for $\psi$ of the form (\ref{HW}) leads automatically
to $\tau$ of the form
\be
\tau=exp\left(-\frac{1}{2}\sum C_{mn}t_nt_m\right)\times\Theta
\left({\cal A}(P_{\infty})+z_0+\sum (\frac{t_j}{2\pi i}(\Omega_{jk})\right)
\label{ttau}
\ee
($C_{mn}=-q_{mn}$) satisfying the corresponding Hirota bilinear identity.
Then as in \cite{cz,ne} one arrives at an asymptotic form $\tau=
exp[(F/\epsilon^2)+O(1/\epsilon)]$ with the quadratic part of $F$ equal
to $(1/2)\sum q_{mn}T_nT_m$.  Hence one obtains $F_{mn}\sim q_{mn}$ and a
Riemann surface version of dKP corresponds to the Whitham equations.  Indeed
we note that from $d\Omega_n\sim z^{-n}-\sum(q_{mn}/m)z^m\sim
z^{-n}-\sum(F_{mn}/m)z^m$ one has
\be
\partial_k\Omega_n=-\sum_{m,n=1}^{\infty}\frac{F_{mnk}}{m}z^m=
\partial_n\Omega_k=-\sum_{m,n=1}^{\infty}\frac{F_{mkn}}{m}z^m
\label{Whit}
\ee


\newpage

\begin{thebibliography}{cc}
%
\bibitem{ab} M. Adler and P. vanMoerbeke,
Comm. Math. Phys., 147 (1992), 25-56; Comm. Pure Appl. Math., 47 (1994), 5-37
%
\bibitem{al} L. Martinez Alonso and E. Moreno,
Chaos, solitons, and fractals, 5 (1995), 2213-2237
%
\bibitem{af} L. Alvarez Gaum\'e, C. Gomez, G. Moore, and C. Vafa,
Nucl. Phys. B, 305 (1988), 455-521
%
\bibitem{ag} L. Alvarez Gaum\'e and C. Reina,
Strings and superstrings, World Scientific, 1988, pp. 135-216
%
\bibitem{ac} L. Alvarez Gaum\'e, and C. Vafa,
Comm. Math. Phys., 106 (1986), 1-40
%
\bibitem{ae} S. Aoyama and Y. Kodama,
Mod. Phys. Lett. A, 9 (1994), 2481-2492; hep-th 9505122
%
\bibitem{bc} H. Baker,
Abelian functions, Cambridge Univ. Press, 1897
%
\bibitem{bb} E. Belokolos, A. Bobenko, V. Enolskij, A. Its, and V.
Matveev,
Algebro-geometric approach to nonlinear integrable equations, Springer,
1994
%
\bibitem{ca} R. Carroll,
Topics in soliton theory, North-Holland, 1991
%
\bibitem{ch} R. Carroll and Y. Kodama,
Solution of the dispersionless Hirota equations, hep-th 9506007, Jour.
Phys. A, 28 (1995), 6373-6387
%
\bibitem{ci} R. Carroll,
Jour. Nonlin. Sci., 4 (1994), 519-544; Teor. Mat. Fizika, 99 (1994),
220-225
%
\bibitem{cl} R. Carroll,
Proc. NEEDS 94, World Scientific, 1995, pp. 24-33
%
\bibitem{cn} R. Carroll,
solv-int 9511009
%
\bibitem{co} R. Carroll,
solv-int 9606005, Proc. Second World Congress Nonlinear Analysts,
Athens, 1996, North-Holland, to appear
%
\bibitem{cp} R. Carroll,
hep-th 9607219, Mod. Phys. Lett. A, to appear
%
\bibitem{cs} R. Carroll,
hep-th 9610216
%
\bibitem{cz} R. Carroll and J. Chang,
The Whitham equations revisited, to appear
%
\bibitem{cc} R. Carroll and B. Konopelchenko,
Lett. Math. Phys., 28 (1993), 307-319
%
\bibitem{ct} R. Carroll,
Amer. Math. Soc. Contemp. Math., Vol. 122 (1991), pp. 23-28; Proc.
NEEDS Workshop, Dubna, 1990, Skpringer, 1991, pp. 2-5
%
\bibitem{cm} I. Cherednik,
Basic methods of soliton theory, World Scientific, 1996
%
\bibitem{dq} E. D'Hoker, I. Krichever, and D. Phong,
hep-th 9609041, 9609145, and 9610156
%
\bibitem{da} B. Dubrovin and S. Novikov,
Russ. Math. Surv., 44 (1989), 35-124
%
\bibitem{dc} B. Dubrovin,
Russ. Math. Surv., 36 (1981), 11-92
%
\bibitem{dd} B. Dubrovin, I. Krichever, and S. Novikov,
Math. Phys. Rev., 3 (1982), 1-150
%
\bibitem{de} B. Dubrovin and S. Novikov,
Math. Phys. Rev., 9 (1991), 3-136
%
\bibitem{fg} H. Farkas and I. Kra,
Riemann surfaces, Springer, 1992
%
\bibitem{fe} J. Fay,
Theta functions on Riemann surfaces, Springer Lect. Notes Math., 352, 1973
%
\bibitem{fz} O. Forster,
Lectures on Riemann surfaces, Springer, 1992
%
\bibitem{fb}
F. Fucito, A. Gamba, M. Martellini, and O. Ragnisco,
Inter. Jour. Mod. Phys. B, 6 (1992), 2123-2147
%
\bibitem{gh} J. Gibbons and Y. Kodama,
Singular limits of dispersive waves, Plenum, 1994, pp. 61-66
%
\bibitem{ga} P. Griffiths,
Introduction to algebraic curves, Amer. Math. Soc., 1989
%
\bibitem{ge} P. Grinevich, A. Orlov, and E. Schulman,
Important developments in soliton theory, Springer, 1993, pp. 283-301
%
\bibitem{gd} P. Grinevich and A. Orlov,
Problems in modern quantum field theory, Springer, 1989, pp. 86-106
%
\bibitem{gf} P. Grinevich,
Singular limits of dispersive waves, Plenum, 1994, pp. 67-88
%
\bibitem{gi} R. Gunning,
Lectures on Riemann surfaces, Princeton Univ. Press, 1966
%
\bibitem{ia} Y. Imayoshi and M. Taniguchi,
An introduction to Teichm\"uller spaces, Springer, 1992
%
\bibitem{ib} H. Itoyama and A. Morozov,
hep-th 9511126, 9512161, and 9601168
%
\bibitem{ha} N. Hawley and M. Schiffer,
Acta Math., 115 (1966), 199-236
%
\bibitem{hb} D. Hejhal,
Mem. Amer. Math. Soc., 129, 1972
%
\bibitem{ks} N. Kawamoto, Y. Namikawa, A. Tsuchiya, and Y. Yamada,
Comm. Math. Phys., 116 (1988), 247-308
%
\bibitem{kn} Y. Kodama and J. Gibbons,
Proceedings Fourth Workshop on Nonlinear and Turbulent Processes in Physics,
World Scientific, 1990, pp. 166-180
%
\bibitem{km} Y. Kodama,
Lecture Lyon workshop, 1991, unpublished
%
\bibitem{kq} Y. Kodama and J. Gibbons,
Phys. Lett. A, 135 (1989), 167-170
%
\bibitem{kr} Y. Kodama,
Phys. Lett. A, 129 (1988), 223-226; Prog. Theor. Phys., Supp. 94 (1988),
184-194
%
\bibitem{kw} N. Kostov,
Lett. Math. Phys., 17 (1989), 95-108
%
\bibitem{ka} I. Krichever,
Funct. Anal. Prilozh., 22 (1988), 200-213
%
\bibitem{kb} I. Krichever,
Russ. Math. Surv., 44 (1989), 145-225
%
\bibitem{ke} I. Krichever,
Math. Phys. Rev., 9 (1991), 1-103
%
\bibitem{kj} I. Krichever,
Comm. Pure Appl. Math., 47 (1994), 437-475
%
\bibitem{kt} I. Krichever and D. Phong,
hep-th 9604199
%
\bibitem{kf} I. Krichever,
hep-th 9611158
%
\bibitem{mz} M. Matone,
Phys. Lett. B, 357 (1995), 342-348; hep-th 9506181
%
\bibitem{mb} R. Miranda,
Algebraic curves and Riemann surfaces, Amer. Math. Soc., 1995
%
\bibitem{ma} D. Mumford,
Lectures on theta, 1 and 2, Birkh\"auser, 1983-84
%
\bibitem{ne} T. Nakatsu and K. Takasaki,
hep-th 9509162
%
\bibitem{nd}
S. Novikov, S. Manakov, L. Pitaevskij, and V. Zakharov,
Theory of solitons, Plenum, 1984
%
\bibitem{ob} W. Osgood,
Lehrbuch der Funktionentheorie, Chelsea, 1965
%
\bibitem{ra} Yu. Rodin,
The Riemann boundary problem on Riemann surfaces, Reidel, 1988
%
\bibitem{sf} N. Seiberg and E. Witten,
Nucl. Phys. B, 426 (1994), 19-52
%
\bibitem{sc} G. Springer,
Introduction to Riemann surfaces, Chelsea, 1981
%
\bibitem{tb} I. Taimanov,
alg-geom 9609016
%
\bibitem{ta} K. Takasaki and T. Takebe,
Inter. Jour. Mod. Phys. A, Supp. 1992, pp. 889-922; Rev. Math. Phys.,
7 (1995), 743-808
%
\bibitem{ya} A. Yaremchuk,
Inverse Probs., 10 (1994), 957-973
%
\bibitem{za} E. Zverovich,
Uspekhi Mat. Nauk, 26 (1971), 117-192
%




\end{thebibliography}
\end{document}